\documentclass[sigconf]{acmart}
\usepackage{multirow}
\usepackage{algorithmic}
\usepackage{algorithm}
\AtBeginDocument{%
  \providecommand\BibTeX{{%
    \normalfont B\kern-0.5em{\scshape i\kern-0.25em b}\kern-0.8em\TeX}}}

\copyrightyear{2024}
\acmYear{2024}
\setcopyright{acmlicensed}
\acmConference[FPGA '24] {Proceedings of the 2024 ACM/SIGDA International Symposium on Field Programmable Gate Arrays}{March 3--5, 2024}{Monterey, CA, USA.}
\acmBooktitle{Proceedings of the 2024 ACM/SIGDA International Symposium on Field Programmable Gate Arrays (FPGA '24), March 3--5, 2024, Monterey, CA, USA}
\acmPrice{15.00}
\acmISBN{979-8-4007-0418-5/24/03}
\acmDOI{10.1145/XXXXXX.XXXXXX}

\acmSubmissionID{143}

\begin{document}

\title{ Efficient Message Passing Architecture for GCN Training on HBM-based FPGAs with Orthogonal Topology On-Chip Networks}

\author{Qizhe Wu}
\email{wqz1998@mail.ustc.edu.cn}
\affiliation{%
  \institution{University of Science and Technology of China}
  \city{Hefei}
  \state{Anhui}
  \country{China}
}

\author{Letian Zhao}
\email{zhaolt@mail.ustc.edu.cn}
\affiliation{%
  \institution{University of Science and Technology of China}
  \city{Hefei}
  \state{Anhui}
  \country{China}
}

\author{Yuchen Gui}
\email{guiyuchen@mail.ustc.edu.cn}
\affiliation{%
  \institution{University of Science and Technology of China}
  \city{Hefei}
  \state{Anhui}
  \country{China}
}

\author{Huawen Liang}
\email{lhw233@mail.ustc.edu.cn}
\affiliation{%
  \institution{University of Science and Technology of China}
  \city{Hefei}
  \state{Anhui}
  \country{China}
}

\author{Xiaotian Wang}
\email{wxtdsg@mail.ustc.edu.cn}
\affiliation{%
  \institution{University of Science and Technology of China}
  \city{Hefei}
  \state{Anhui}
  \country{China}
}

\author{Xi Jin}
\email{jinxi@ustc.edu.cn}
\affiliation{%
  \institution{University of Science and Technology of China}
  \city{Hefei}
  \state{Anhui}
  \country{China}
}








\renewcommand{\shortauthors}{Trovato and Tobin, et al.}

\begin{abstract}
Graph Convolutional Networks (GCNs) are state-of-the-art
deep learning models for representation learning on graphs. However, the efficient training of GCNs is hampered by constraints in memory capacity and bandwidth, compounded by the irregular data flow that results in communication bottlenecks. 
To address these challenges, we propose a message-passing architecture that leverages NUMA-based memory access properties and employs a parallel multicast routing algorithm based on a 4-D hypercube network within the accelerator for efficient message passing in graphs. Additionally, we have re-engineered the backpropagation algorithm specific to GCNs within our proposed accelerator. This redesign strategically mitigates the memory demands prevalent during the training phase and diminishes the computational overhead associated with the transposition of extensive matrices. Compared to the state-of-the-art HP-GNN architecture we achieved a performance improvement of $1.03\times \sim 1.81\times$.
\end{abstract}




\received{20 February 2007}
\received[revised]{12 March 2009}
\received[accepted]{5 June 2009}

\maketitle

\section{Introduction}
GCNs serve as a leading deep learning for graph and have found advanced applications in recommendation systems \cite{zhu2019aligraph}, transportation networks \cite{derrow2021eta}, and drug discovery. Graph-based deep learning has become the mainstream model, fueling the field's rapid growth.

With the proliferation of applications and increased graphs, training graph neural networks have become inefficient and face several challenges. Firstly, unlike inference, training requires retaining the output from different layers during forward to calculate backpropagation. However, single-machine accelerators, such as GPU and FPGA, cannot provide sufficient out-chip storage and bandwidth for training as the graph increases. To mitigate this issue, GraphSage \cite{hamilton2017inductive}, GraphSAINT \cite{zeng2019graphsaint}, VRGCN \cite{chen2017stochastic}, and FastGCN \cite{chen2018fastgcn} have introduced sampling-based mini-batch training methods make training large graphs feasible under this computation paradigm. 

Secondly, GCN is sensitive to the order of execution. Previous work on inference accelerators \cite{geng2020awb,zhou2021blockgnn, zhang2021boostgcn,liang2020engn, li2021gcnax} has explored this issue. However, in training, backpropagation is performed in the reverse order of forward, so optimizing the computing order of forward propagation only benefits some models. Additionally, during the training process, there is a significant need for matrix transposition, which adds complexity to optimizing the dataflow. It requires a holistic consideration of computation and storage \cite{guirado2021characterizing}. 

Thirdly, high communication overhead caused by frequent node transmission. In single-machine setups, a centralized shared Unified Memory Architecture (UMA) structure is commonly used in the Symmetric Multi-Processing (SMP) model. However, during graph aggregation, on-chip computing units generate multiple random accesses in the shared memory, leading to tough competition and longer latency for local computing resources \cite{yang2022drgn,tian2022fp,li2021sgcnax}. Resulting in lower average utilization of computing resources \cite{abadal2021computing, yuan2022qegcn, wang2021gnnadvisor}. Addressing these challenges requires efficient communication schemes and memory management strategies in GNN training.

Fourthly, High-bandwidth memory (HBM) plays a crucial role in facilitating ample bandwidth for neural network training on FPGAs. In particular, HBM has been shown to provide significant bandwidth \cite{wissolik2017virtex,holzinger2021fast}, surpassing traditional DDR4 channels by a magnitude on FPGA platforms \cite{hu2021graphlily,kang2022fpga}. However, when applying Graph Neural Network (GNN) computations, we encounter diverse memory access patterns, with the combination phase involving sequential address access and extended burst lengths, while the aggregation phase exhibits random memory access with shorter burst lengths. This poses a notable challenge in implementing a scalable GNN training architecture on an HBM-based FPGA system. We need to mitigating bandwidth reduction caused by multiple AXI ports accessing the same pseudo-channel during aggregation.

Our main contributions are as follows: 

(1) We propose a dedicated Multi-Cores GNN training accelerator for high-performance message passing on HBM-based FPGAs with NUMA-based memory access. The on-chip networks of the accelerator adopt a strictly orthogonal hypercube topology, and design a highly concurrent routing mechanism for GNN applications.

(2) We redesigned the dataflow of GNN backpropagation in the FPGA. It can decrease the off-chip memory requirements during training while reducing additional matrix transpose.

(3) We evaluate the performance on Xilinx UltraScale+ VCU128. In comparison to the state-of-the-art GNN training architecture HP-GNN\cite{lin2022hp}, we achieved a improvement of 1.03$\times$ $\sim$  1.81$\times$.
\begin{figure*} [htbp]
  \flushleft 
  \includegraphics[scale=0.185]{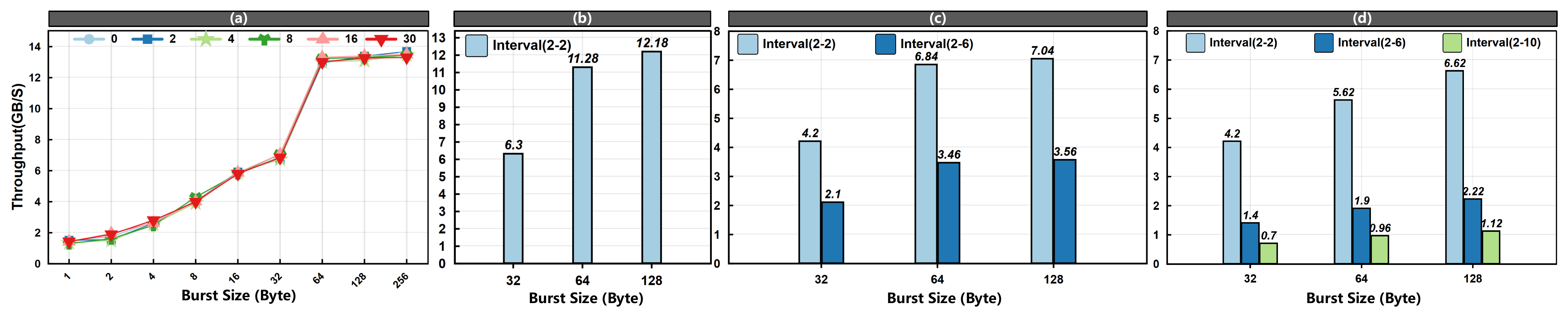}
  \caption{HBM bandwidth benchmarking. (a) AXI port reads for local channels. (b)(c)(d) Concurrent reads for non-local channels.}
  \label{ch}
\end{figure*}

\section{Background}
$\mathcal{G}=(\mathcal{V,E}$) represents an undirected graph, whereas $\mathcal{V}=(v_{1} ,...,v_{n})$ and $\mathcal{E}=( e_{1} ,...,e_{k})$ represent the set with $n$ nodes and $k$ edges, respectively. $X\in \mathbb{R}^{n\times d}$ represents the feature matrices of all nodes, each of which has a feature vector of length $d$. The structure of the graph is given by the adjacency matrix $A\in \mathbb{R}^{n\times n}$, where $A_{( i,j)} =1$ if $( v_{i} ,v_{j}) \in \mathcal{E}$  else $A_{( i,j)} =0$. The purpose of this study is to obtain, through training, a low-latitude vector representation $x_{v}$ of all $ v\in \mathcal{V}$ so that it can more accurately predict node classification. GNNs utilize the neighborhood aggregation approach to achieve this representation. Specifically, GNNs update the representation of a node by aggregating the representations of its neighbors in a recursive manner. Formally, the $ l^{th}$ layer of a GNN can be expressed as $x_{v}^{( l+1)} =\mathbf{UPDATE}\left( x_{v}^{( l)} ,\bigoplus _{u\in \mathcal{N}( v)}\mathbf{MSG}\left( x_{u}^{( l)} ,x_{v}^{( l)}\right)\right)$ \cite{gilmer2017neural}, where $ x_{v}^{( l)}$ represents node $v$ at the $ l^{th}$ layer. $\mathcal{N}(v)$ represents node $v$'s neighbors, excluding $v$ itself. For node $v$, the message function $\mathbf{MSG}\left(\cdot\right)$ calculates messages from its neighbors. These messages are aggregated using a permutation-invariant aggregation function. Then $\mathbf{UPDATE}( \cdot )$ transforms the aggregated features at $v$.

Our work is based on the GCN \cite{welling2016semi} layer and defines it as a matrix multiplication expression \cite{fey2019fast},\cite{fey2021gnnautoscale} in the form of Equ.~\ref{eq1}, $X^{(l)}$ is the node embedding matrix at the $l^{th}$ layer, $W^{(l)}$ is the weight matrix of the $l^{th}$ layer, $\tilde{A} =\tilde{D}^{-\frac{1}{2}}A\tilde{D}^{-\frac{1}{2}}$ is the normalized adjacency matrix, where $\tilde{D}$ is the degree matrix of $A+I$. Since $ \tilde{A} $ is often sparse and feature $ X^{(l)} $ is typically dense, the GCN layer is expressed as the expression of General Matrix Multiply (GEMM) and Sparse-Dense Matrix Multiplication (SPMM) are abbreviated in the equations as $\mathbf{GM}$ and $\mathbf{SM}$, $ \sigma $ illustrate of nonlinear activation functions.
\begin{equation} \label{eq1}
     X^{( l+1)} = \sigma \left(\mathbf{SM}\left(\tilde{A} ,\mathbf{GM}\left( X^{( l)} ,W^{( l)}\right)\right)\right)
\end{equation}

As shown by Equ.~\ref{eq2}, the partial differentiation of the loss function $ \mathcal{L}$ with respect to feature matrix $X(l)$ is error matrix of the $l^{th}$ layer, whereas $ \sigma '$ is the partial differentiation of the nonlinear activation function. According to Equ.~\ref{eq3}, the gradient matrix $G^{(l)}$ of the $l^{th}$ layer is the partial differential of the $ \mathcal{L}$ to weight matrix $W^{(l)}$.

\begin{equation} \label{eq2}
E^{( l)} =\frac{\partial \mathcal{L}}{\partial X^{( l)}}  =\mathbf{GM}\left(\mathbf{SM}\left(\tilde{A}^{T} , \sigma '\left( E^{( l+1)}\right)\right) ,\left( W^{( l)}\right)^{T}\right)
\end{equation}

\begin{equation} \label{eq3}
G^{( l)} =\frac{\partial \mathcal{L}}{\partial W^{( l)}} =\mathbf{GM}\left(\left( X^{(  l)}\right)^{T}\mathbf{,SM}\left(\tilde{A}^{T} , \sigma '\left( E^{( l+1)}\right)\right)\right)
\end{equation}

Special attention should be given that when there is a change in the computation sequence during the forward (prioritizing GM or SM), corresponding adjustments in the computation sequence need to be made in the context of backward and gradient computation.

In the parameter update phase of the optimizer take Stochastic Gradient Descent (SGD) as an example, which is an iterative method for optimizing an objective function with suitable smoothness properties. Choose an initial parameters $W^{(l)}$ and learning rate $ \eta$, see Equ.~\ref{eq4}, repeat until $ \mathcal{L}$ is obtained an approximate minimum.

\begin{equation} \label{eq4}
W_{t+1}^{( l)} =W_{t}^{( l)} -\eta \nabla \mathcal{L}\left( X,W_{t}^{( l)}\right)   
\end{equation}

\begin{figure*} [htbp]
  \flushleft
  \includegraphics[scale=0.212]{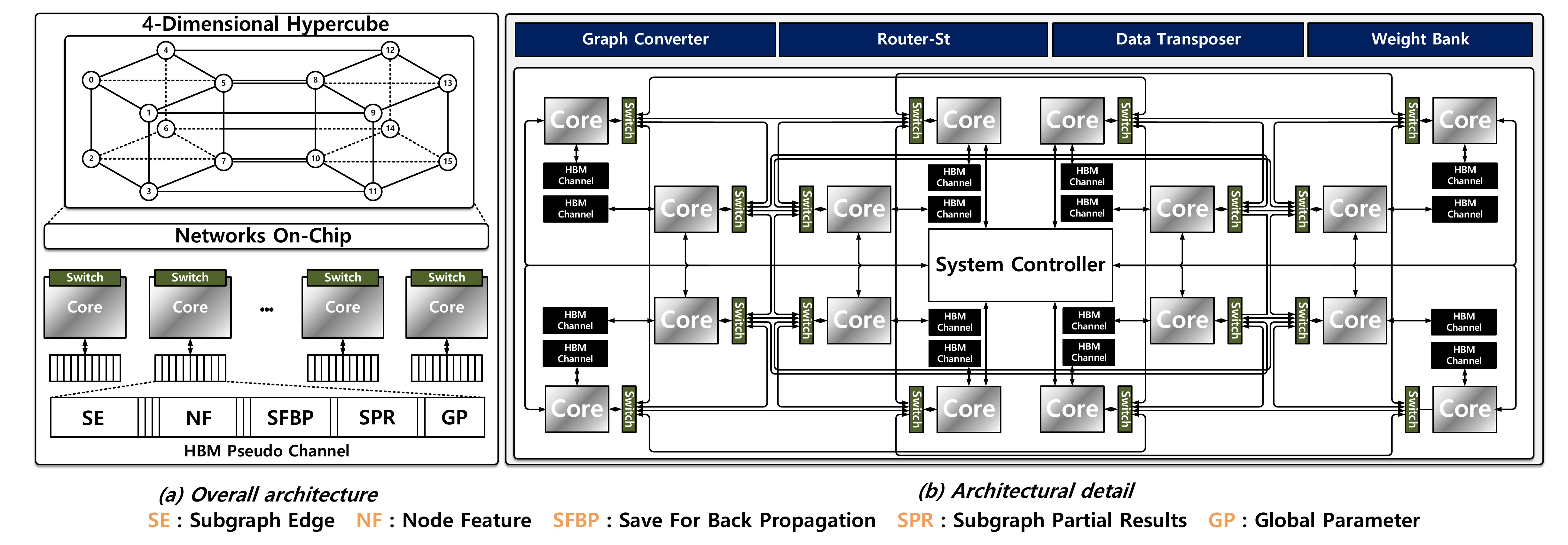}
  \caption{Overall architecture of GNN Training.}
  \label{overal}
\end{figure*}
\section{motivation}
Due to the high bandwidth advantage of HBM, larger-scale GNN computations can now be deployed on FPGAs. The combination phase in GNN involves sequential address access with larger burst lengths, while the aggregation phase requires random access with smaller burst lengths. We analyze and test the access behaviors that may occur in GNN computations with HBM. In Fig.\ref{ch}(a), we present a fundamental bandwidth test where we examine the read bandwidth of AXI interfaces using local pseudo-channels labeled 0, 2, 4, 8, 16, and 30, under various burst lengths.

Fig.\ref{ch}(b), (c), and (d) analyze the bandwidth when non-local AXI interfaces generate multiple memory access requests to the same HBM pseudo-channel. In GNN, the computation engine may generate multiple memory access requests to the same HBM pseudo-channel during aggregation. These requests can come from local AXI interfaces or arbitrary channel AXI interfaces. In Fig.\ref{ch}(b), we test the case where two access requests are issued by AXI interfaces at intervals of two pseudo-channel lengths from the target channel. Compared to using local AXI interfaces, the read bandwidth decreases by 13.7$\%$ and  6.8$\%$  with a burst length of 64 and 128. Fig.\ref{ch}(c) tests the scenario where four access requests are simultaneously generated for a single channel, spaced at intervals of 2 and 6 channel lengths from the target channel. Each channel generates two memory access requests. The read bandwidth drops by 21.1$\%$ and 19.6$\%$ with a burst length of 64 and 128. Lastly, in Figure \ref{ch}(d), we test the situation where six access requests are simultaneously generated for a single channel, spaced at intervals of 2, 6, and 10 channel lengths from the target channel. The read bandwidth decreases by 35.1$\%$ and 24.4$\%$ with a burst length of 64 and 128. 

\textbf{Conclusion}: the number of concurrent accesses to the target channel and the interface distance from which requests are issued significantly impact HBM bandwidth, making it challenging to fully leverage the high bandwidth of HBM during aggregation. As computational resources increase, bandwidth gradually becomes the bottleneck. This leads to lower utilization rates, reducing the scalability of the accelerator and limiting the performance of HBM.

\section{Architecture}
This chapter covers the overall architecture of the FPGA in section 4.1, core architecture in section 4.2, the on-chip routing algorithm and router design for parallel multicast message passing of graphs in aggregation phase on the FPGA in section 4.3, the operator splitting and the dataflow optimization in GNN training in section 4.4. 

\subsection{Architecture Overview}
Fig.\ref{overal}(a) illustrates the decentralized GCN training architecture, dispersing data and functionality across multiple computation nodes. Our approach to operator partitioning and model deployment aims to maximize communication between cores and minimize communication density between HBM channels.

The architecture utilizes the NUMA characteristics of the DSM model, deploying 16 computational cores with non-global addressing mode in HBM. Each core exclusively occupies 2 HBM pseudo-channels, with no direct access to non-corresponding channels. Each HBM channel stores data dependencies for task deployment, including node features (NF), subgraph edges (SE), save for backpropagation (SFBP), subgraph partially results (SPR), and global parameters (GP). This approach provides high bandwidth for GNN combination phase while shifting aggregation workload from HBM to on-chip interconnect network.

Fig.\ref{overal}(b) demonstrates the architecture in detail, employing a 4-D hypercube topology between cores. Dedicated street routers (Router-St) facilitate message passing between cores. An efficient data transfer algorithm tailored to GNN models in orthogonal topology on-chip interconnect networks is designed to address insufficient bandwidth for HBM inter-channel communication. Practical deployment is based on the typical structure of the hypercube in orthogonal topology networks.

During training, edge aggregation follows row-major order in the forward pass and column-major order in backpropagation. To avoid redundant storage of edges, we store the adjacency matrix in COO format and use a Graph Converter to switch between row-major and column-major orders. A dense matrix transposer (Data Transposer) is designed to transpose matrices needed for backpropagation. The Weight Bank synchronizes global weights after updates, and the system controller periodically uses it to update global parameters within HBM pseudo-channels.

\subsection{Core Architecture}
\begin{figure} [htbp]
  \flushleft
  \includegraphics[scale=0.47]{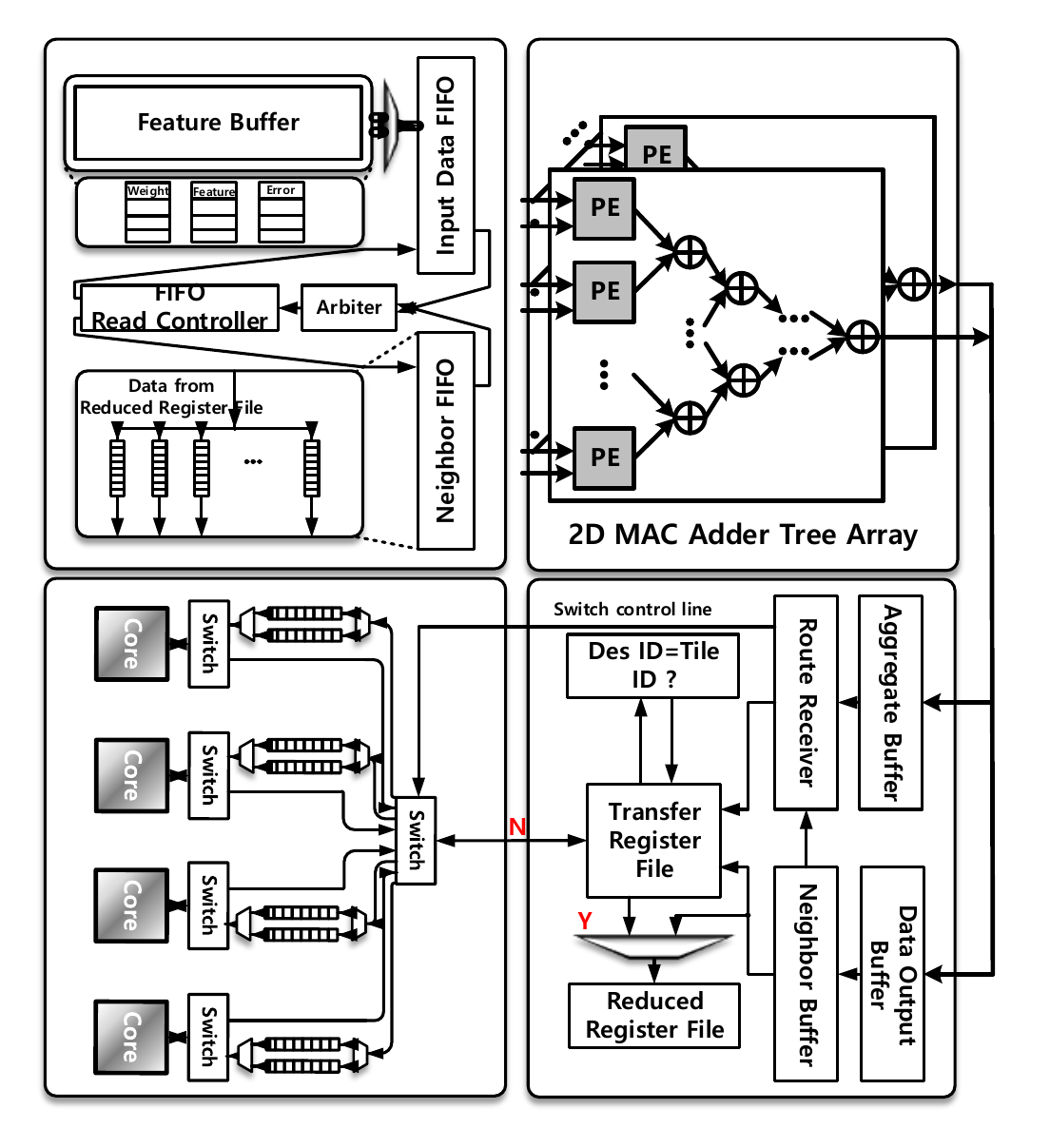}
  \caption{Core architecture.}
  \label{core}
\end{figure}
Each core utilizing a 2D MAC Adder Tree for privode computing power (Fig.\ref{core}). Matrix multiplication, vector, or scalar operations are implemented by MAC direct output or adder tree output. The Computation to Communication (CTC) ratio is enhanced by decomposing the training process into fusion computation processes of tensors and graphs. In the forward, the feature combination task is initially assigned to each core. Features and partially updated weights are written to the Feature Buffer of each core for block matrix multiplication. Additionally, the Output Buffer and Neighbor Buffer are prepared to facilitate Ping-Pong operation.

After completing the current phase of partial combination tasks, the results are written to the Neighbor Buffer for message passing and the Aggregate Buffer for storing the aggregated features. The PE array then initiates the next phase of feature combination tasks, with the Output Buffer storing temporary output results. The on-chip network between each core is triggered for message passing.

The routing instructions are decoded by the Route Receiver. System Controller distributes the routing instruction to the internal regions of every core in each cycle. Based on the routing instruction, the Route Receiver reads the feature from the Neighbor Buffer and stores it in the Transfer Register File. \textbf{If multiple neighbors are within the locally targeted core, they are locally aggregated before message transmission.} These neighbors are locked in the Reduced Register File and written to the Neighbor FIFO.

The Arbiter within the core determines which FIFO the PE array should read from based on the empty/full status of the Neighbor FIFO. When aggregation is required, it switches the data path from the Input Data FIFO to the Neighbor FIFO. Each core continuously switches between matrix tasks and scalar multiplication and accumulation tasks.

After that, the PE array writes the results for routing (routing can only begin after merging all messages within each core). Once the data is ready for routing, cores initiates street routing. When data passes through the Transfer Register File of each core, the Route Receiver decodes the current routing instruction to determine the physical channel for dispatching the data in the next phase.

When the packet reaches the Transfer Register File of its destination core, the features from the Aggregate Buffer and the node features from the Transfer Register File can be written into the Reduced Register File for aggregation in the Neighbor FIFO. After alculations, the results are written back to the original address within the local core's Aggregate Buffer. Each calculation stage needs to write the results in the Aggregate Buffer back to memory as the output of the graph convolution layers before initiating the next stage of the pipeline.

\begin{figure} [htbp]
  \flushleft
  \includegraphics[scale=0.09]{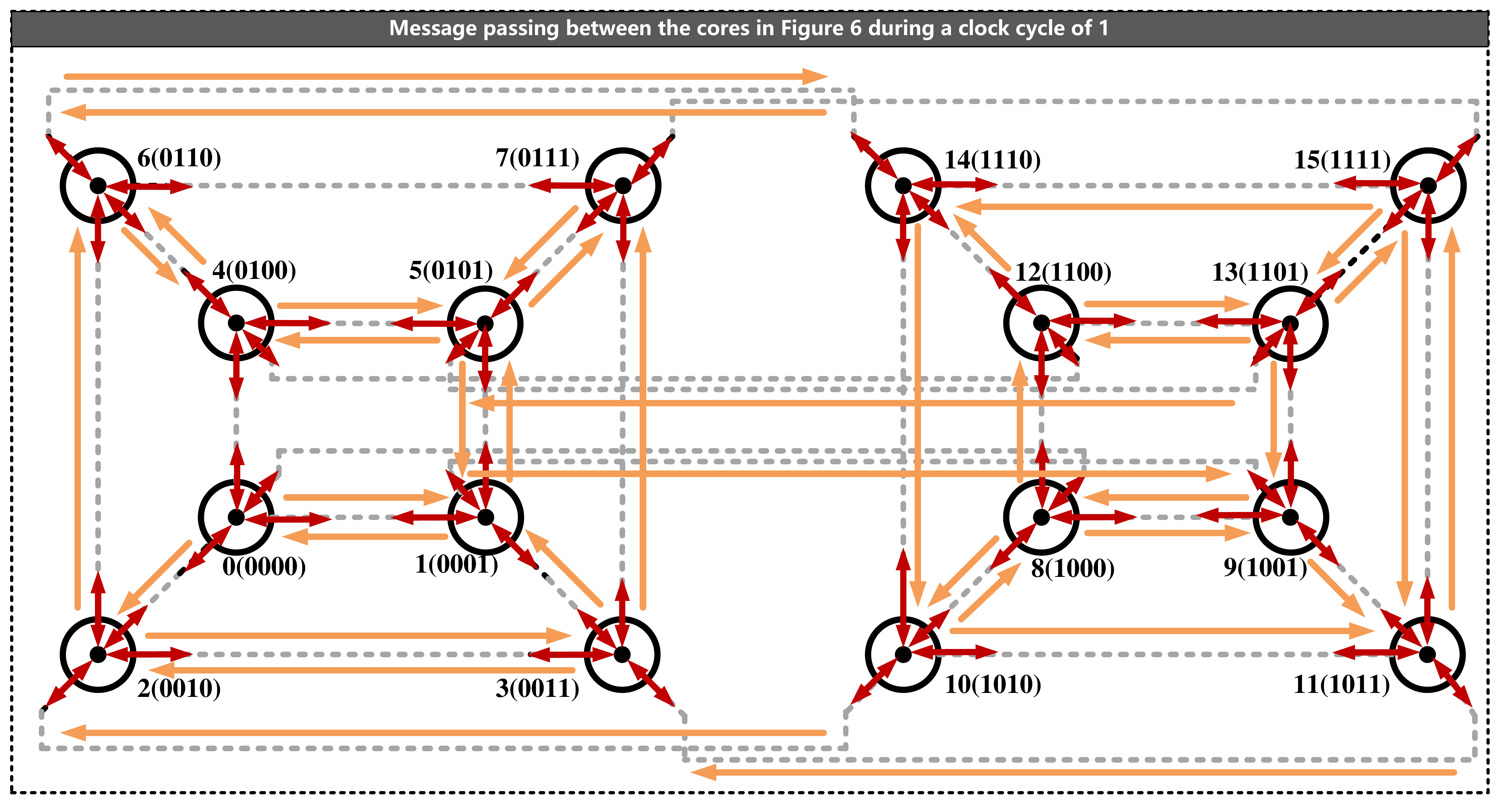}
  \caption{4-D hypercube on-chip network.}
  \label{hypercube}
\end{figure}

\subsection{High-performance On-chip Aggregation}

\subsubsection{Orthogonal Topology On-chip Network}
The main advantage of strict orthogonal topology is the routing mechanism, which allows the implementation of efficient routing algorithms in hardware. We can calculate the shortest distance between two computing nodes by the offsets in each dimension. Every computing node is defined by an n-dimensional binary coordinate $(x_{n-1}, x_{n-2},..., x_1, x_0)$. The necessary condition for two adjacent computing nodes is that there should be a $j$ such that $y_j = x_j \pm 1$, and for any $i \neq j$, and $0 \leq i \leq n-1$, $y_i = x_i$.

We implement a 4-D hypercube topology with strict orthogonality as the interconnection medium between cores (Fig.\ref{hypercube}). The network comprises 16 computing nodes, allowing bidirectional transmission to neighboring nodes in the same cycle. For the network topology to be orthogonal, the nodes must be arranged in an n-dimensional space, with each bit of binary code representing a dimension in the strict orthogonal topology. Consequently, the arrangement of links in each dimension must create an offset.

\subsubsection{Switch Model and Routing Rule}
\begin{figure} [htbp]
  \flushleft
  \includegraphics[scale=0.12]{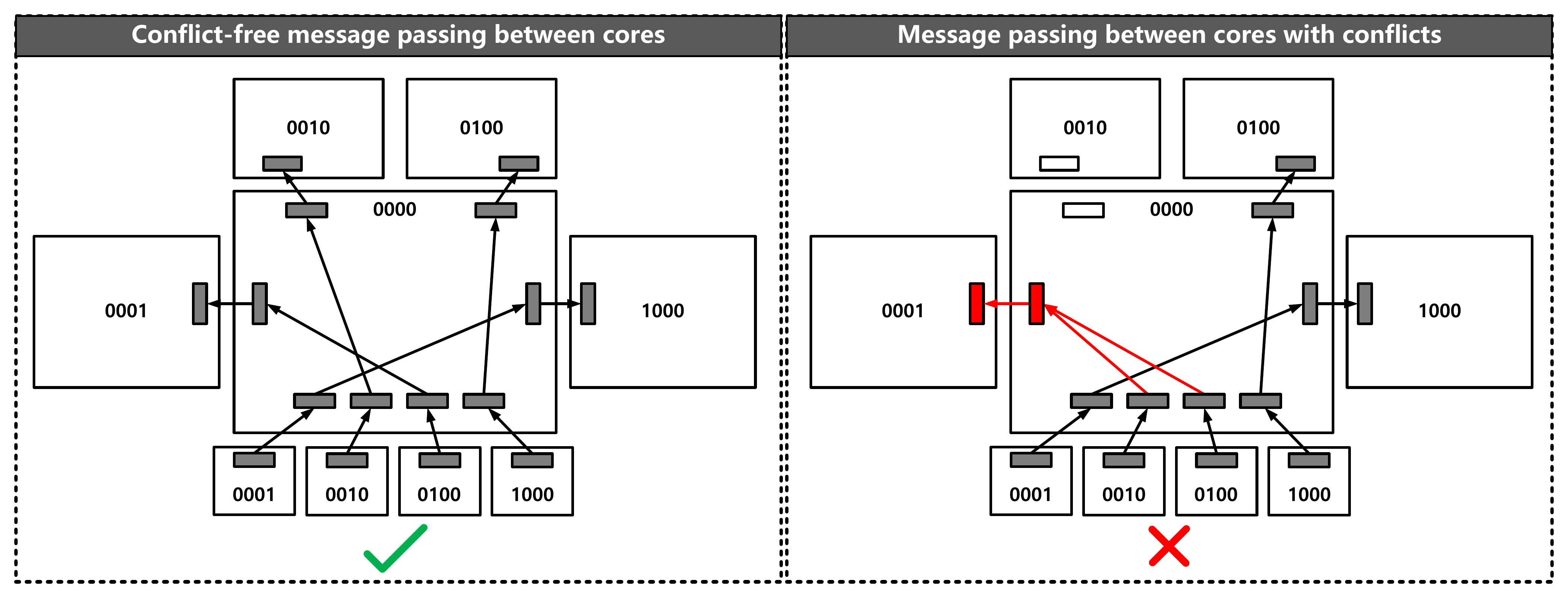}
  \caption{Switch model in each core.}
  \label{switch}
\end{figure}

To maximize throughput (Fig.\ref{switch}), where two data transmission lines are present between adjacent computing nodes (one for sending and one for receiving). To prevent the need for increasing the bit width of interconnect lines, in a single clock cycle, each node can receive a maximum of one message from each of its four connected cores (maximum receive limit per core is 4). However, from the perspective of the sending core, if multiple messages attempt to use the same output channel simultaneously, it will lead to a deadlock.

\begin{figure*} [htbp]
  \flushleft
  \includegraphics[scale=0.083]{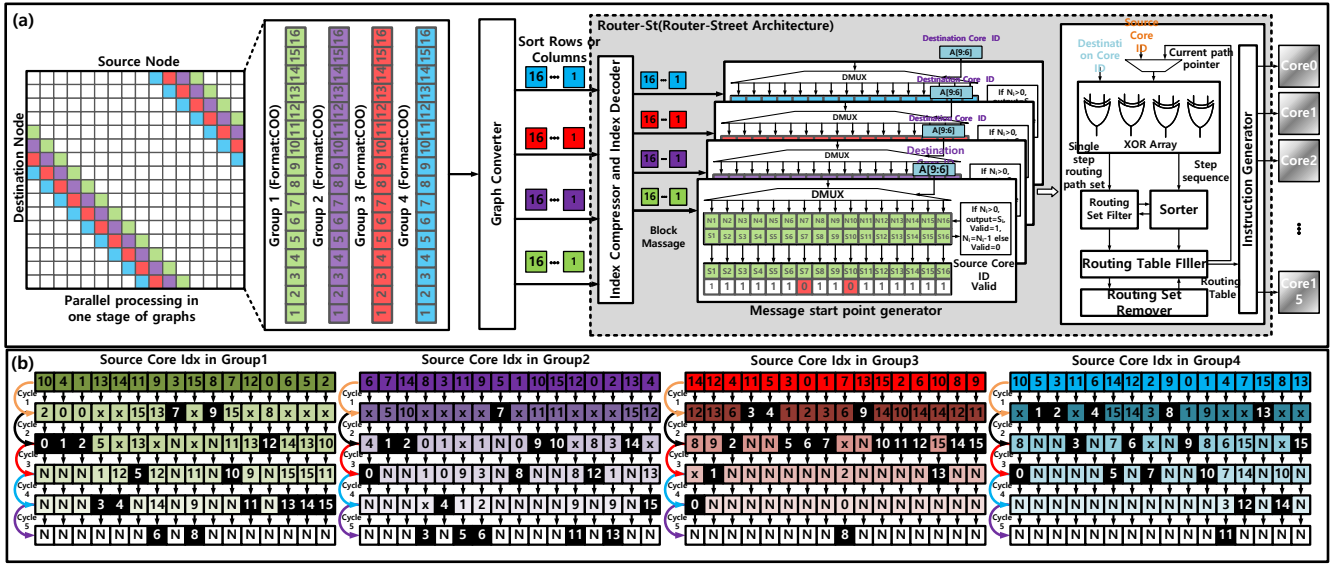}
  \caption{The processing of graphs and the Router-St architecture.}
  \label{rst}
\end{figure*}

\subsubsection{Parallel Multicast Routing Strategy}

To maintain high transmission bandwidth while minimizing costs to the system's on-chip interconnects, we have developed three optimizations based on the graph's characteristics. \textbf{Firstly, we employ a path selection mechanism} to calculate the shortest one-step path for message transmission, reducing latency and distance on on-chip interconnects. \textbf{Secondly, we utilize data compression} to merge and compress neighboring nodes during the graph routing. \textbf{Lastly, we leverage transmission parallelism} to achieve a deadlock-free 4-group parallel multicast, \textbf{where 16 nodes simultaneously send four messages to other nodes.}

For efficient processing, we partition the large adjacency matrix into smaller blocks that are independently handled by each core. Each core can handle a graph of up to 1024 nodes, further divided into 16 parts as shown in Fig.\ref{rst}(a). We store the adjacency matrix using a "diagonal storage" method. It offers the advantage of storing only the upper or lower triangular portion, reducing storage requirements for undirected graph training.

To enable parallel aggregation processing, we partition the 1024-node subgraph into multiple stages. Each stage processes four diagonals, equivalent to 64 blocks, with each group containing 16 block queues. The COO formatted graph for each block undergoes sorting of rows or columns based on the current computational stage and node aggregation direction, using the Graph Converter module. The sorted data is then dispatched to the Router-St, which consists of four main sections: \textbf{(1) graph compression and message encoding/decoding, (2) message starting point generator, (3) route calculation, and (4) route instruction generator}.

\begin{figure} [htbp]
  \flushleft
  \includegraphics[scale=0.09]{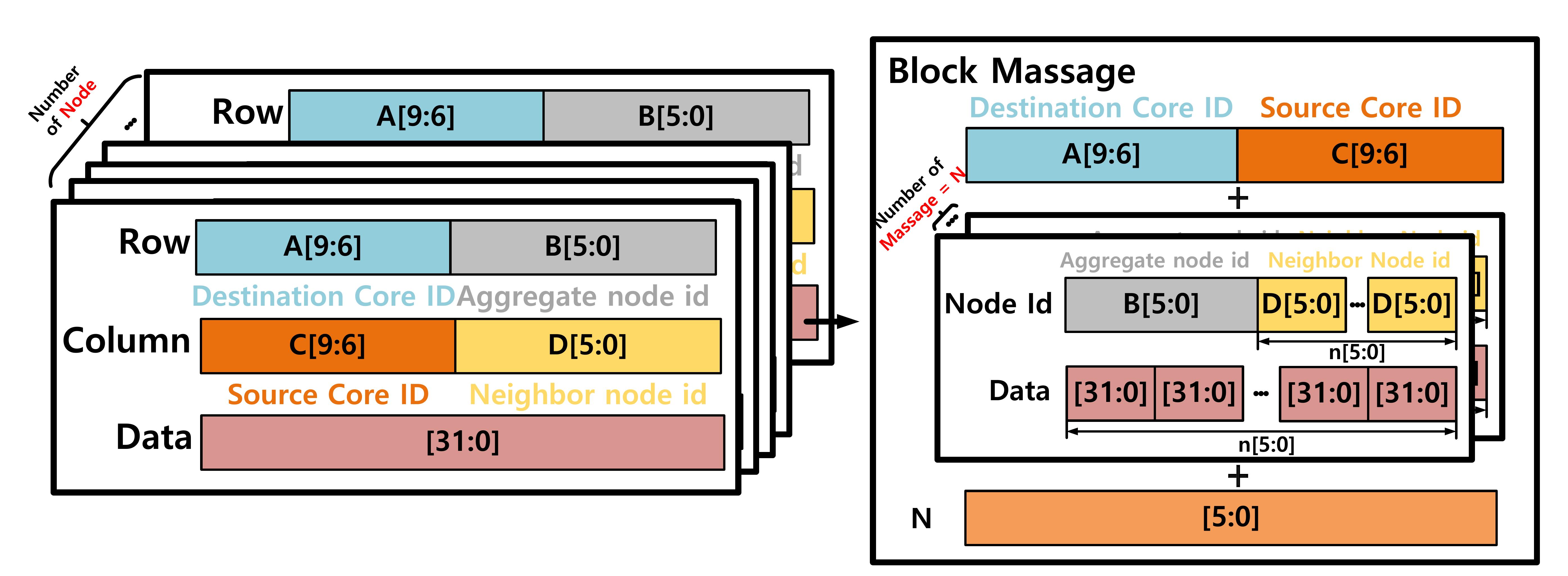}
  \caption{The sparse matrix of the graph is converted into a compressed block message representation.}
  \label{bm}
\end{figure}

\textbf{Index Compressor and Index Decoder:} compression of the graph data of 64 blocks distributed in 4 groups (blue, red, purple, and green regions in Fig.\ref{rst}) into Block Messages (BM) in COO format (Fig.\ref{bm}). During forward propagation, nodes are aggregated row-wise, resulting in even distribution and execution of 1024 nodes among 16 cores. Each core's buffer stores the feature of 64 nodes.

For a 1024-node adjacency matrix, the column index's high 4 bits decode as the core id, representing the current source core id of the node. The lower 6 bits of the column index indicate the node's address in the source core Neighbor Buffer, specifying the location of neighboring nodes in the accelerator. The high 4 bits of the row index correspond to the destination core id, while the remaining 6 bits (B: Aggregate node id) represent the base address of the master node stored in the destination core's Aggregate Buffer.

Within a block (subgraph of 64 nodes) in a group, all nodes share the same A (Destination Core ID) and C (Source Core ID) encoding. By traversing the Aggregate node ID (B) of all nodes in a block, nodes with matching Aggregate node IDs are combined into a single message expression. This means that in core A, neighbors of node B are located in the D row of the Neighbor Buffer in core C. The adjacency matrix in a block is compressed into a Block Message of the form A + C + N, where N represents the number of times A and C need to communicate. The Block Message is stored in the buffer of the corresponding source core id. When a core receives a start routing instruction, messages scheduled for transmission are combined by matching Neighbor Node ID (D) with the same Aggregate node ID (B) within a core before sending.

\begin{algorithm}[!h]
    \caption{Parallel Multicast Routing}
    \label{alg:AOA}
    \renewcommand{\algorithmicrequire}{\textbf{Input:}}
    \renewcommand{\algorithmicensure}{\textbf{Output:}}
    \begin{algorithmic}[1]
        \REQUIRE A$\left[p\right]$:Source index vector A has p element; B$\left[p\right]$:De-\newline stination index vector B has p element 
        \ENSURE Routing\_table:2-D table with p columns 
        \STATE $Path\_Set,Step\_Seq\leftarrow XOR\_Array (A,B);$ 
        \WHILE{zero\_all$(Step\_Seq)$}
         \STATE $index\_step\leftarrow argsort(Step\_Seq);$ 
         \STATE $Path\_Set\leftarrow Set\_Filter(Path\_Set);$ 
            \STATE $cycle\_router\_path \leftarrow Initial(p)$
            \FOR{each $i \in [1,p-1]$}
            \IF{$Path\_Set\big[index\_step\left[i\right]\big] \neq \emptyset$   \& \newline $Step\_Seq\big[index\_step\left[i\right]\big]>0$}
                \STATE $path\_id \leftarrow Rand\_sel(Path\_Set\big[index\_step\left[i\right]\big]);$
                \\ \STATE $cycle\_router\_path\big[index\_step\left[i\right]\big]\leftarrow path\_id;$
                \\ \STATE $Path\_Set \leftarrow delet\_set(Path\_Set,path\_id,index\_ \newline step[i],Routing\_point);$
            \ELSE
                \STATE $cycle\_router\_path\big[index\_step\left[i\right]\big]\leftarrow \times;$
            \ENDIF
        \ENDFOR
        \STATE $Routing\_table.insert(cycle\_router\_path);$
        \STATE $Routing\_point \leftarrow Generate\_rp(Routing\_table,cycl$-
        $e\_router\_path,Routing\_point);$
        \STATE $Path\_Set,Step\_Seq\leftarrow XOR\_Array (Routing\_point,\newline B);$
        \emph{\scriptsize{\#Update the next path and step}}
        \ENDWHILE
        \RETURN Routing\_table
    \end{algorithmic}
\end{algorithm}

\textbf{Message Start Point Generator:} to enhance transmission parallelism, we aim to utilize starting points from 16 different cores for the messages. Within the same group, the source core id encoding of 16 Block Messages fulfills this requirement, as does the destination core id. Hence, we extract A, C, and N from all Block Messages to generate routing table, then sorting of Block Messages in ascending order based on destination core id. Each column represents the number of messages sent from a specific core (S1-S16) to the corresponding core (0-15). With each generation of a starting vector, its corresponding N is decremented. According to the switch model, the theoretical maximum number of messages that a 4-D hypercube network can send in one cycle is 64. Thus, a maximum of four messages can originate from the same core in one cycle. To avoid conflicts, we expand the starting point vector into a one-dimensional vector, ensuring that each number occurs a maximum of four times. As depicted in Fig.\ref{rst}(b) as example, we parallelize message transmission for the four groups by extracting a source core id vector from each. In this process, each number within the same group is unique, and no number appears more than four times.

\textbf{Routing computation:} involves the XOR Array, Routing Set Filter, Sorter, Routing Table Filler, and Routing Set Remover. These modules work together to maintain deadlock-free routing calculations on the generated starting vector. The Routing Table Filler generates the routing table (Fig.\ref{rst}(b), to demonstrate the results of routing computation as example, arrow in the Fig.\ref{hypercube} indicates the routing of cycle 1).

\begin{figure} 
  \centering
  \includegraphics[scale=0.15]{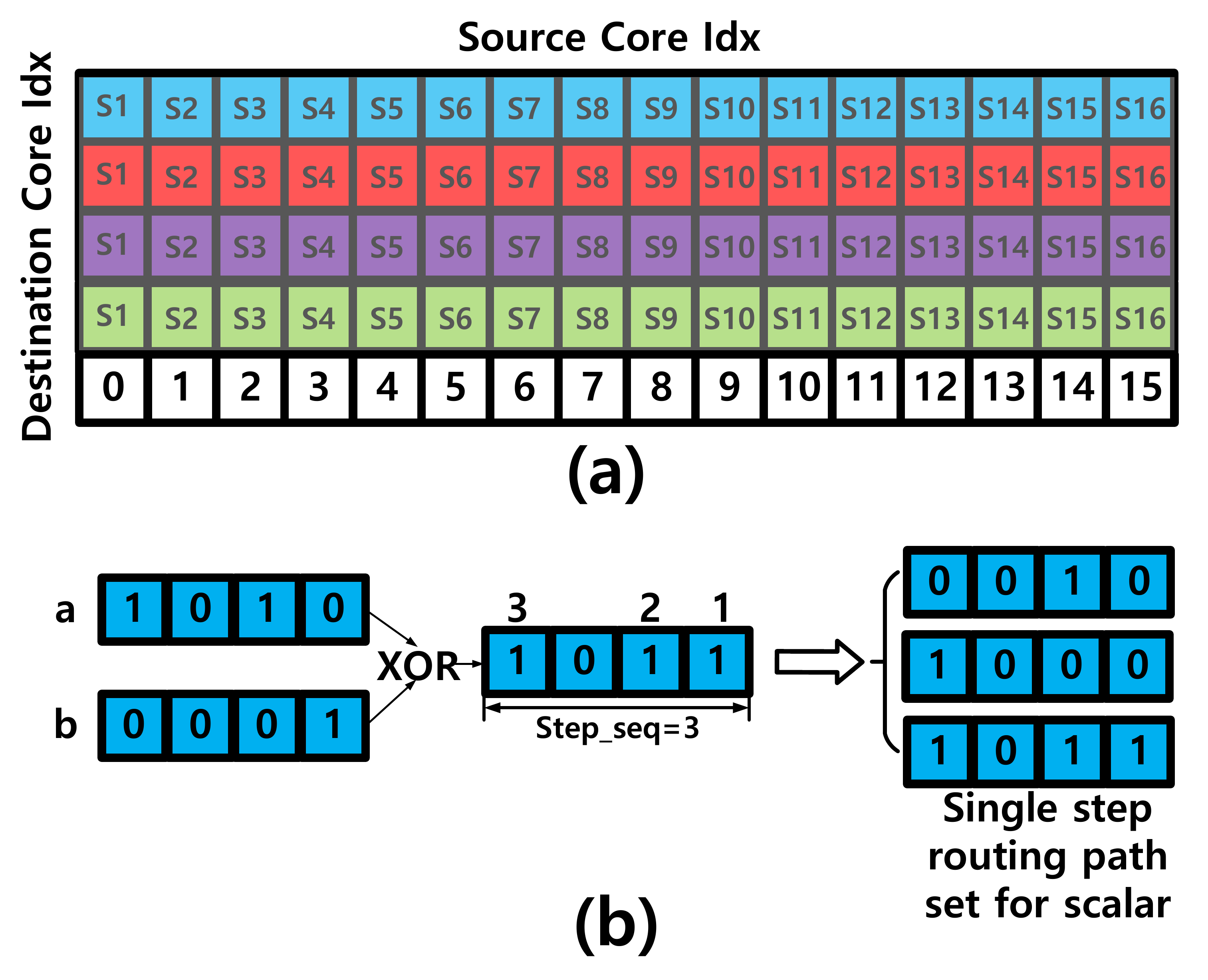}
  \caption{Routing computation.}
  \label{rc}
\end{figure}

\textbf{XOR Array(Algorithm \ref{alg:AOA} line 1):} the module receives the starting point vector (length of 64) for the first row of all groups in Fig.\ref{rst}(b) line 1, along with the corresponding destination core vector (white region in Fig.\ref{rc}(a)). It then generates a single-step path set for the given vectors. A calculation example is shown in Fig.\ref{rc}(b), demonstrating the process of obtaining the single-step path set between scalar variables "a" and "b". The XOR operation on "a" and "b" highlights the differences in the resulting output's bit positions. To determine the single-step path set of "A", the values of the corresponding bit positions in "a" (with an XOR result of 1) need to be negated. The step length represents the number of binary digits with a value of 1 in the XOR result, indicating the shortest time in cycles required for messages to be transmitted from "A" to "B".

\textbf{Routing Set Filter (Algorithm \ref{alg:AOA} line 4):} filters the single-step path set generated by the XOR Array. \textbf{Constraint 1: Each receiving core can only receive a maximum of 4 messages simultaneously.} Therefore, the single-step path set of all nodes will be scanned, and any number that appears more than four times will be removed. The priority of removing the elements in the node set is determined by sorting the path length (elements of node sets with more alternative paths will be removed first). This is a dynamic process, and the priority queue must be adjusted after each removal to ensure that the path set length is balanced. After filtering, most nodes should have one or two alternative paths.

\textbf{Sorter (Algorithm \ref{alg:AOA} line 3):} responsible for sorting the step sequence queue generated by the XOR Array. Specifically, it groups the node numbers with the same step length and outputs the corresponding indices. \textbf{Routing Table Filler:} \textbf{(Algorithm \ref{alg:AOA} line 8-9)} fills the deadlock-free routing channels of the framework according to the indices from Sorter. The nodes with shorter step lengths are given priority, as they can reach the destination faster and release the channel for other nodes to use. Please note that the nodes with longer step lengths may have more alternative paths and, therefore, have a lower priority. Moreover, the filler selects one of the single-step paths available in the node's collection randomly as the target node for the next cycle. 
\begin{table*}[htbp]
  \centering
\scriptsize
  \caption{Theoretical analysis of time and storage complexity for training GCN layers with various execution orders. Assuming the current layer is the $kth$ layer from the bottom of the network. Here, $b$ denotes the batch size, $n$ denotes the number of $k-1$ hop neighbors in the input batch, $d$ denotes the length of the input node feature vector, and $\overline{n}$ represents the number of 1-hop neighbors of $n$, where $X\in \mathbb{R}^{\overline{n} \times d}$. $A$ is the corresponding adjacency matrix, where $A\in \mathbb{R}^{n \times \overline{n}}$, $e$ represents the non-zero elements in $A$,  where $W \in \mathbb{R}^{d \times h}$, $c$ represents the total number of classes for node classification, and $E$ and $E^{\mathcal{L}}$ represent the errors in backpropagation of the $(k+1)th$ layer and the loss function layer, respectively.}
    \label{table_fx}%
  \setlength{\tabcolsep}{4.7mm}{
  \renewcommand{\arraystretch}{1.4}
    \begin{tabular}{cccccccccccccccccccccccccccccccccccccccccccccccccccccccc}
    &&&&&&&&\\
    \midrule
    \midrule
         \multicolumn{4}{c}{\textbf{Stage}} & \multicolumn{18}{c}{\textbf{Forward}}  & \multicolumn{12}{c}{\textbf{Backward}} & \multicolumn{12}{c}{\textbf{Gradient}} \\
     \midrule
     \midrule
     \multicolumn{4}{c}{\textbf{Operation}} & \multicolumn{12}{c}{\textbf{GM-SM}} & \multicolumn{6}{c}{\textbf{Transpose}} & \multicolumn{12}{c}{\textbf{SM-GM}} & \multicolumn{6}{c}{\textbf{GM}} & \multicolumn{6}{c}{\textbf{Transpose}} \\
    \midrule
     \midrule
    \multirow{3}*{\textbf{CoAg}} & \multicolumn{3}{c}{Math}  
    & \multicolumn{12}{c}{$A(XW)$} 
    & \multicolumn{6}{c}{$A^{T},W^{T}$}    
    & \multicolumn{12}{c}{\textcolor{red}{$\left( A^{T}E\right)W^{T}$}}
    & \multicolumn{6}{c}{\textcolor{red}{$X^{T}\left( A^{T} E\right)$}}  
    & \multicolumn{6}{c}{\textcolor{red}{$X^{T}$}}\\
    \cline{2-46}
 & \multicolumn{3}{c}{Time Complexity} & \multicolumn{12}{c}{$O(\overline{n}dh) +O(eh)$}  & \multicolumn{6}{c}{$O(\overline{n}e)+O(hd)$} & \multicolumn{12}{c}{$O(eh)+O(\overline{n}dh)$} & \multicolumn{6}{c}{$O(\overline{n}dh)$}  & \multicolumn{6}{c}{$O(\overline{n}d)$} \\
 \cline{2-46}
& \multicolumn{3}{c}{Storage Complexity} & \multicolumn{12}{c}{$O(\overline{n}d)+O(\overline{n}h)+O(e)$} & \multicolumn{6}{c}{$O(e)$}  & \multicolumn{12}{c}{$O(\overline{n}h)+O(nh)$}  & \multicolumn{6}{c}{$\backslash$} & \multicolumn{6}{c}{$O(\overline{n}d)$}\\
\midrule
\midrule
     \multicolumn{4}{c}{\textbf{Operation}} & \multicolumn{12}{c}{\textbf{SM-GM}} & \multicolumn{6}{c}{\textbf{Transpose}} & \multicolumn{12}{c}{\textbf{GM-SM}} & \multicolumn{6}{c}{\textbf{GM}} & \multicolumn{6}{c}{\textbf{Transpose}} \\
    \midrule
     \midrule
    \multirow{3}*{\textbf{AgCo}} & \multicolumn{3}{c}{Math} 
    & \multicolumn{12}{c}{$(AX)W$}
    & \multicolumn{6}{c}{$A^{T},W^{T}$}    
    & \multicolumn{12}{c}{\textcolor{magenta}{$ A^{T}\left(EW^{T}\right)$}} 
    & \multicolumn{6}{c}{\textcolor{magenta}{$(AX)^{T}E$}} 
    & \multicolumn{6}{c}{\textcolor{magenta}{$(AX)^{T}$}}\\
    \cline{2-46}
 & \multicolumn{3}{c}{Time Complexity} & \multicolumn{12}{c}{$O(ed)+O(ndh)$}  & \multicolumn{6}{c}{$O(\overline{n}e)+O(hd)$}
 & \multicolumn{12}{c}{$O(ndh)+O(ed)$} 
 & \multicolumn{6}{c}{$O(ndh)$}  
 & \multicolumn{6}{c}{$O(nd)$} \\
 \cline{2-46}
& \multicolumn{3}{c}{Storage Complexity}
& \multicolumn{12}{c}{$O(\overline{n}d)+O(nd)+O(e)$}
& \multicolumn{6}{c}{$O(e)$} 
& \multicolumn{12}{c}{$O(nd)+O(nh)$} 
& \multicolumn{6}{c}{$\backslash$}
& \multicolumn{6}{c}{$O(nd)$}\\
\midrule
\midrule

     \multicolumn{4}{c}{\textbf{Operation}} & \multicolumn{12}{c}{\textbf{GM-SM}} & \multicolumn{6}{c}{\textbf{Transpose}} & \multicolumn{12}{c}{\textbf{SM-GM}} & \multicolumn{6}{c}{\textbf{GM}} & \multicolumn{6}{c}{\textbf{Transpose}} \\
    \midrule
     \midrule
    \multirow{3}*{\textbf{Ours CoAg}} & \multicolumn{3}{c}{Math} 
    & \multicolumn{12}{c}{$A(XW)$}
    & \multicolumn{6}{c}{$W^{T}$}    
    & \multicolumn{12}{c}{\textcolor{red}{$W\left(E^{T}A\right)$}} 
    & \multicolumn{6}{c}{\textcolor{red}{$\left(E^{T}A\right)X$}}
    & \multicolumn{6}{c}{\textcolor{red}{$(E^{\mathcal{L}})^{T}$}}\\
    \cline{2-46}
 & \multicolumn{3}{c}{Time Complexity} 
 & \multicolumn{12}{c}{$O(\overline{n}dh) +O(eh)$}  
 & \multicolumn{6}{c}{$O(hd)$}
 & \multicolumn{12}{c}{$O(eh)+O(\overline{n}dh)$}
 & \multicolumn{6}{c}{$O(\overline{n}dh)$} 
 & \multicolumn{6}{c}{$O(bc)$} \\
 \cline{2-46}
& \multicolumn{3}{c}{Storage Complexity} & \multicolumn{12}{c}{$O(\overline{n}d)+O(\overline{n}h)+O(e)$} & \multicolumn{6}{c}{$\backslash$}  & \multicolumn{12}{c}{$O(\overline{n}h)+O(nh)$}  & \multicolumn{6}{c}{$\backslash$} & \multicolumn{6}{c}{$\backslash$}\\

\midrule
\midrule
     \multicolumn{4}{c}{\textbf{Operation}} & \multicolumn{12}{c}{\textbf{SM-GM}} & \multicolumn{6}{c}{\textbf{Transpose}} & \multicolumn{12}{c}{\textbf{GM-SM}} & \multicolumn{6}{c}{\textbf{GM}} & \multicolumn{6}{c}{\textbf{Transpose}} \\
    \midrule
     \midrule
    \multirow{3}*{\textbf{Ours AgCo}} & \multicolumn{3}{c}{Math} 
    & \multicolumn{12}{c}{$(AX)W$}
    & \multicolumn{6}{c}{$W^{T}$}    
    & \multicolumn{12}{c}{\textcolor{magenta}{$(W(E^{T}))A$}}
    & \multicolumn{6}{c}{\textcolor{magenta}{$E^{T}(AX)$}} 
    & \multicolumn{6}{c}{\textcolor{magenta}{$(E^{\mathcal{L}})^{T}$}}\\
    \cline{2-46}
 & \multicolumn{3}{c}{Time Complexity} & \multicolumn{12}{c}{$O(ed)+O(ndh)$}  & \multicolumn{6}{c}{$O(hd)$}
 & \multicolumn{12}{c}{$O(ndh)+O(ed)$} 
 & \multicolumn{6}{c}{$O(ndh)$}  
 & \multicolumn{6}{c}{$O(bc)$} \\
 \cline{2-46}
& \multicolumn{3}{c}{Storage Complexity}
& \multicolumn{12}{c}{$O(\overline{n}d)+O(nd)+O(e)$}
& \multicolumn{6}{c}{$\backslash$} 
& \multicolumn{12}{c}{$O(nd)+O(nh)$} 
& \multicolumn{6}{c}{$\backslash$}
& \multicolumn{6}{c}{$\backslash$}\\

     \midrule
\midrule
    \end{tabular}}%
\end{table*}%

\textbf{Constraint 2: The recipient cannot receive two or more messages simultaneously from the same core id.} To ensure that the proposed framework satisfies Constraint 2, the \textbf{Routing Set Remover (Algorithm \ref{alg:AOA} line 10)} is employed. After one node is filled, the Remover traverses the path collection of other nodes and removes any node that does not meet Constraint 2. 

Encountering an empty set during the filling process can occur due to path removal by the Routing Set Filter or the Routing Set Remover in order to prevent routing deadlock. To handle empty sets, we label the path with an "$\times$" and temporarily store it in a virtual channel until its destination is determined in the next cycle's routing filling process. After filling in the vectors, we generate the current path pointers for all groups (Algorithm \ref{alg:AOA}, line 16). In cycle 1, the address pointer vector of Group 1 is (2,0,0,13,14,15,13,7,15,9,15,12,8,6,5,2) as shown in Fig.\ref{rst}(b). We then input the address pointer vectors of all groups into the XOR Array to calculate and fill the path collection for the next cycle accordingly. For example, in the 12th column of Group 1 in Fig.\ref{rst}(b), the data is inserted into the virtual channel buffer leading to core 13 during the first cycle, starting at core id 12. core 13 receives data from core 12 in the second cycle, and each core selects the current data source based on the command, either the real channel buffer or the virtual channel buffer.

This process continues until the calculation of the Routing Table is completed, which happens when the offset of the current address pointer vector input into the Routing Table matches the step offset of the destination vector. This process enables the generation of a highly parallel message-passing network on a 4-D hypercube. In our hardware architecture, it is possible to send up to 64 messages in just four cycles at the fastest.


\textbf{Instruction Generator:} produce routing instructions for each core. Each routing instruction is comprised of 25 bits. The Head bit helps determine if the instruction is a routing table header. If it is, each core must read the corresponding Block Message of the Destination ID and merge them locally, that is, on the sending core. Routing starts when the router receives a signal from all cores indicating that merging is complete. The four bits of the Receive Signal specify which message channel each core will open during the current cycle. Send ID specifies that the message to be received is to be delivered to the storage channel corresponding to the ID of the core. Open Channel specifies the ID of the sending channel that is presently open and whether the related data is from a virtual or real channel buffer. Destination ID specifies the ultimate destination of this message. The transmitted data packet format includes 518 bits (Feature: 512bit,Aggregate Node id: 6bit), which consist of the merged node feature vector and a 6-bit Aggregate node ID.


\subsection{Training Dataflow on FPGA}


It has previously been observed in the GCN inference accelerator 
 \cite{geng2020awb, geng2021gcn} that combination-aggregation (CoAg) is computationally less expensive than aggregation- combination (AgCo) \cite{zhang2021boostgcn}. However, this claim does not hold valid during training tasks. This is because training GCN is a semi-supervised learning task, where the size of the adjacency matrix in each layer depends on the batch size and the number of sampled nodes. In training, the adjacency matrix is often rectangular. Hence, executing aggregation first during the process of training can reduce the dimensionality of the node feature matrix, just like performing a combination operation. Therefore, in training process the optimal execution sequence of AgCo and CoAg is often closely related to the properties of the graph and the selection of hyperparameters (batch size and the number of sampling at each stage) during training. Hence, it is necessary to flexibly choose the execution order based on the characteristics of the dataset.	

 We have incorporated a sequence estimator within the system controller to determine the final training order. The algorithm for the estimator is detailed in Table 1 (notations are explained within the comments). The estimator analyzes the complexity of memory and computational operations required throughout the entire computation flow. Before initiating the calculations, we need to configure the hyperparameters of the dataset into registers within the system controller, including batch size ($b$), the number of $k-1$ hop to 1 hop neighbors $n$ in the input batch, the length ($d$) of the input node feature vector, and non-zero elements $e$ in A. Subsequently, the optimal execution order is determined based on the overall computational complexity.

 Due to the high cost of operating HBM on FPGA and the limited memory space, we found that during the computation of gradients in backpropagation, it is necessary to utilize the transpose of the feature matrix corresponding to the forward pass. As a result, both CoAg and AgCo in Table \ref{table_fx} require the storage of the transpose of $X$ or $AX$ for gradient computation. To ensure that it does not affect performance, these calculations need to be precomputed and stored in HBM before backpropagation. This implies that additional HBM storage space is needed during the forward pass for $X^{T}$ or ${AX}^{T}$, and this space can only be released after the computation corresponding to $X$ or $AX$ is completed. This not only places further demands on HBM memory capacity during training but also adds to the cost of memory address mapping and memory access.

 To address this, we optimized the execution order of backpropagation, and the main differences are highlighted in red and pink in Table \ref{table_fx}. It is evident that we begin backpropagation by computing the transpose of the loss function error $E^{\mathcal{L}}$. This is due to the gradual reduction in feature dimensions as we progress through network layers, making the cost of transposing $E^{\mathcal{L}}$ lower. Then, the entire backpropagation is carried out in a transposed form. As a result, what originally required $X^{T}$ now only needs $X$ for computation, reducing memory requirements, memory operation complexity, and minimizing the computational load of transposition. The findings reveal that our computational sequence has a lower time and storage complexity (TC and SC) than AgCo and CoAg. Presented below is a contrast of various operation sequences:

\begin{equation} \label{eqtccoag}
TC(CoAg-Ours CoAg)=O(\overline{n}(e+d))-O(bc)>0
\end{equation}
\begin{equation} \label{eqtcagco}
TC(AgCo- Ours AgCo)=O(\overline{n}e+nd)-O(bc)>0
\end{equation}
\begin{equation} \label{eqsccoag}
SC(CoAg- Ours CoAg)=O(e)+O(\overline{n}d)>0
\end{equation}
\begin{equation} \label{eqscagco}
SC(AgCo- Ours AgCo)=O(e)+O(\overline{n}d)>0
\end{equation}

\section{EXPERIMENTS}

\subsection{Experimental Setup}

We evaluate on a Xilinx Virtex UltraScale+ HBM VCU128 FPGA hosted by a 24-core Xeon (Platinum 8260 @2.4GHz, hyper-threaded). CPU-FPGA communication is via PCIe 3.0 x16 (peak bandwidth: 15.8GB/s). The GraphSAGE neighbor sampler (NS) is used for the mini-batch training. In all experiments, the neighbor sampling size of the 2-hop neighbors as 10 and 25 for 1-hop neighbors respectively, and the hidden dimension is set to 256. We measure the GNN training performance of two-layer GCN model and two-layer GraphSAGE model on one medium-scale graph datasets (Flickr \cite{ zeng2019graphsaint}) and three large-scale graph datasets (Reddit \cite{hamilton2017inductive}, Yelp \cite{ zeng2019graphsaint} and AmazonProducts \cite{ zeng2019graphsaint}). GPU baseline are implemented using the state-of-the-art graph learning framework Pytorch-Geometric (PyG) on A100. The resource utilization of the accelerator on FPGA is depicted in Table \ref{resources}. Within each computational core, the PE array comprises 256 floating-point multiplication units in the TF32 format and 256 floating-point accumulation units in the FP32 format. The entire system operates at 250MHz.

\begin{figure} [htbp]
  \flushleft
  \includegraphics[scale=0.19]{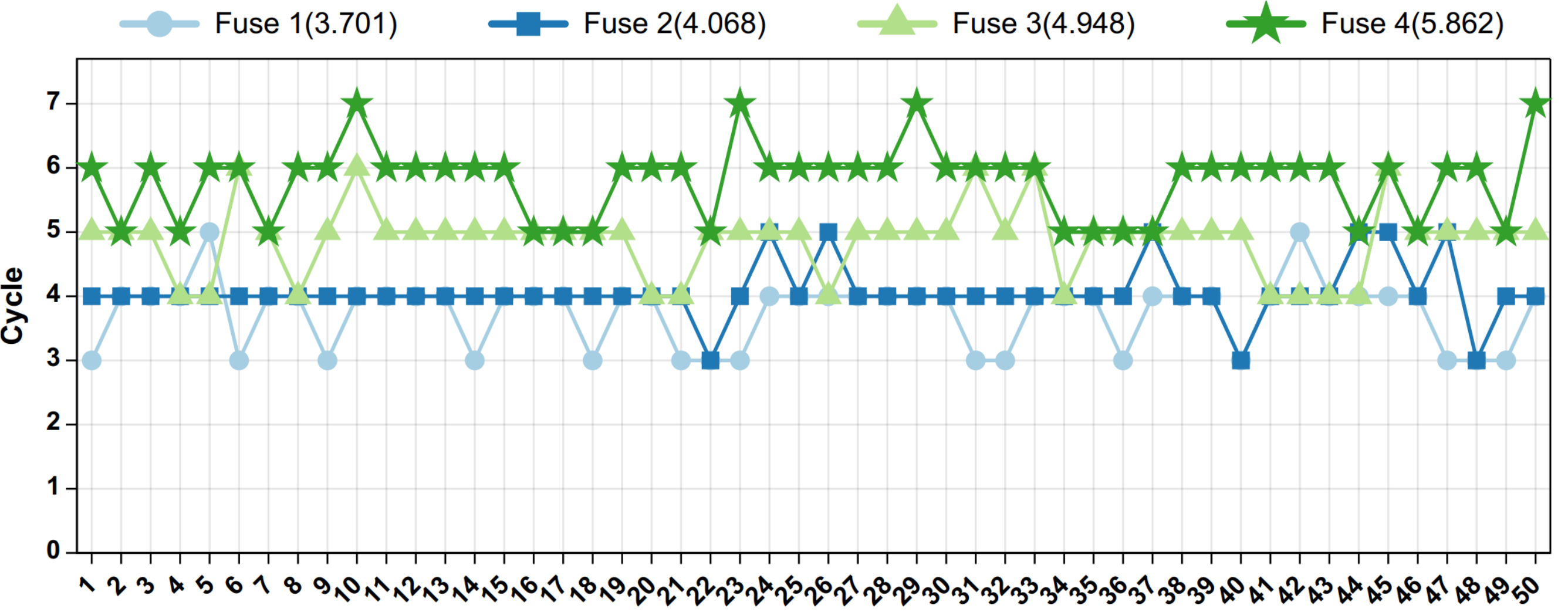}
  \caption{The routing cycle under a random test.}
  \label{random_r}
\end{figure}



\subsection{Routing Performance Analysis}

We tested the route-sending pattern of Router-St under multiple messaging parallelism, as shown in Fig.\ref{random_r}. Fuse1 represents sending 16 messages in parallel (one group in Fig.\ref{rst}(b)). We randomized the starting point vector within each group, creating a random sequence from 0 to 15, and sent each column to different target nodes. Fuse4 represents sending 64 messages in parallel (four groups in Fig.\ref{rst}(b)). We conducted 1000 sets of random starting point stimuli for each Fuse, but only 50 samples are shown in Fig.\ref{random_r}). The parentheses indicate the average receiving cycle of 1000 sets of random sequences. The experiment indicated that our routing mechanism adds only one cycle to the total cycle as the messaging passing increases by one group from Fuse 2 to Fuse 4. In testing, it can provide an aggregate bandwidth of up to 2.96TB/s (${64 \times 4 \times 16 \times \left(\frac{{64}}{{4}}\right)}{{\times \frac{{1}}{{20.13 \times 10^9}}}}$ Byte, here, 64 byte is the transmission bit width of a single data line, 16 corresponds to the computing core, and 4 is the maximum number of messages sent to the interconnected computing nodes in parallel each time. We assume that each of the first four messages is compressed by 16 (64/4) messages in the buffer, and the average routing clock period is 20.13ns). The actual bandwidth provided by the on-chip network without local compression aggregation is 189.4GB/s.

\begin{figure*} 
  \centering
  \includegraphics[scale=0.26]{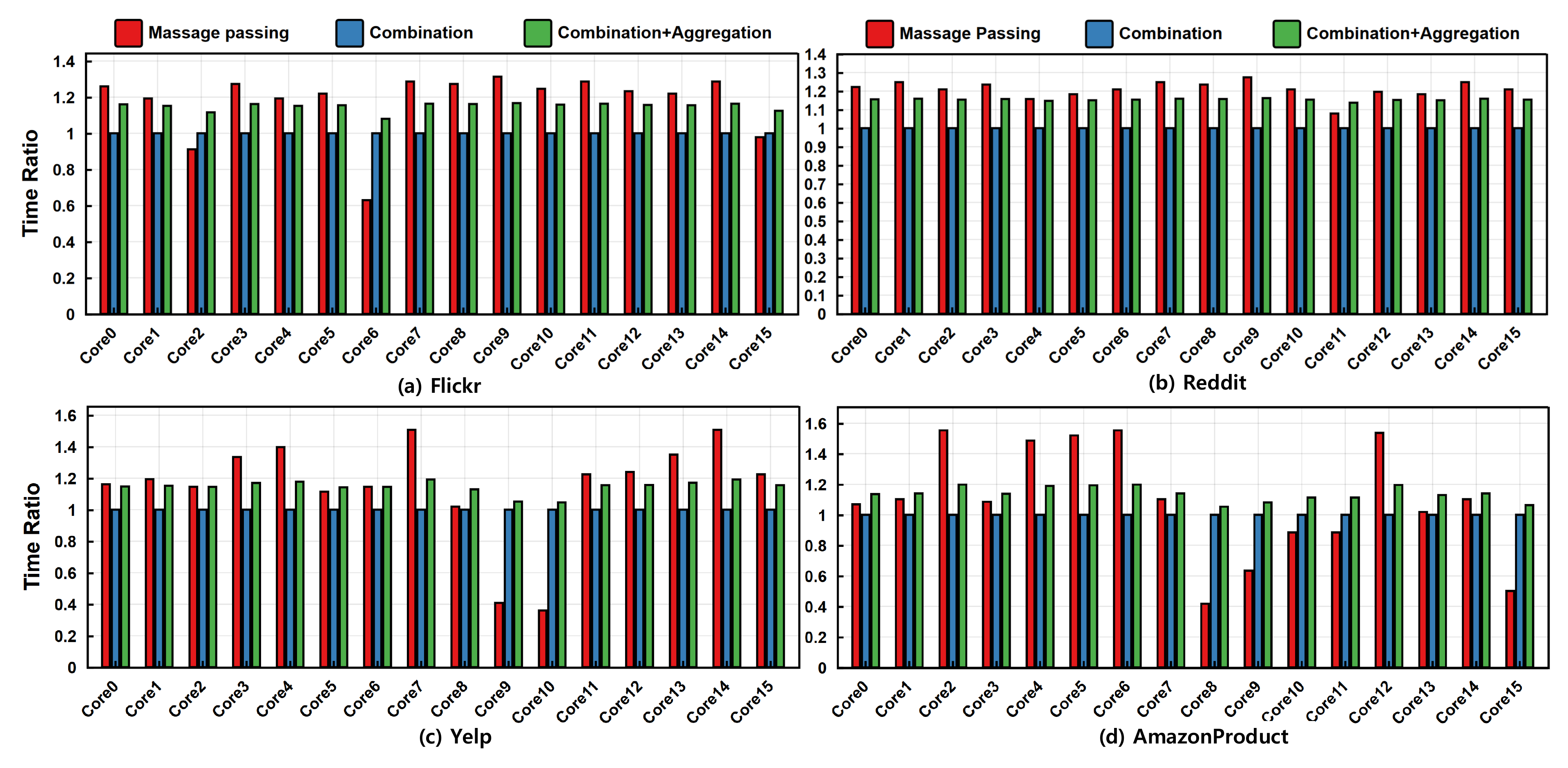}
  \caption{The ratio of message passing time to combination and aggregation.}
  \label{test}
\end{figure*}
\subsection{Hardware Performance Analysis}
\subsubsection{Key Points for Performance Improvement}
In GNN computations, the combination stage involves dense matrix multiplication, while the aggregation stage involves vector multiplication and accumulation. The read bandwidth for the combination stage and the write bandwidth for retaining the forward feature map for backpropagation are provided by HBM. Due to the nature of contiguous accesses, the utilization of HBM channels can be maximized. On the other hand, the bandwidth for the aggregation stage is provided by the on-chip network. 

For a single core scenario, in a GCN layer, when the bandwidth in the combination stage is not the bottleneck, the actual computation time depends on the maximum value between the message passing time and the MAC computation time ($t_{combination}$ + $t_{aggregation}$). The highest utilization is achieved when the MAC computation time is greater than the message passing time, allowing the communication time to be completely hidden. Otherwise, we need to reduce the difference between these two times to improve performance as following equation:
\begin{equation}
    t_{singlecore_{gcn}}=max(t_{massage passing},t_{combation}+ t_{aggregation})
\end{equation}

As shown in Fig.\ref{test}(a)(b)(c)(d), for the majority of cases within a single core, we have shortened the ratio between message passing time and MAC array computation time, thanks to the high concurrency characteristics of the routing algorithm. On average, for the Flickr dataset, the computation-to-communication ratio for each core is 1:1.02. For the Reddit dataset, the average computation-to-communication ratio for each core is 1:1.05. For the Yelp dataset, the average computation-to-communication ratio for each core is 1:0.99. And for the Amazon dataset, the average computation-to-communication ratio for each core is 1:0.94.

\begin{figure} 
  \centering
  \includegraphics[scale=0.26]{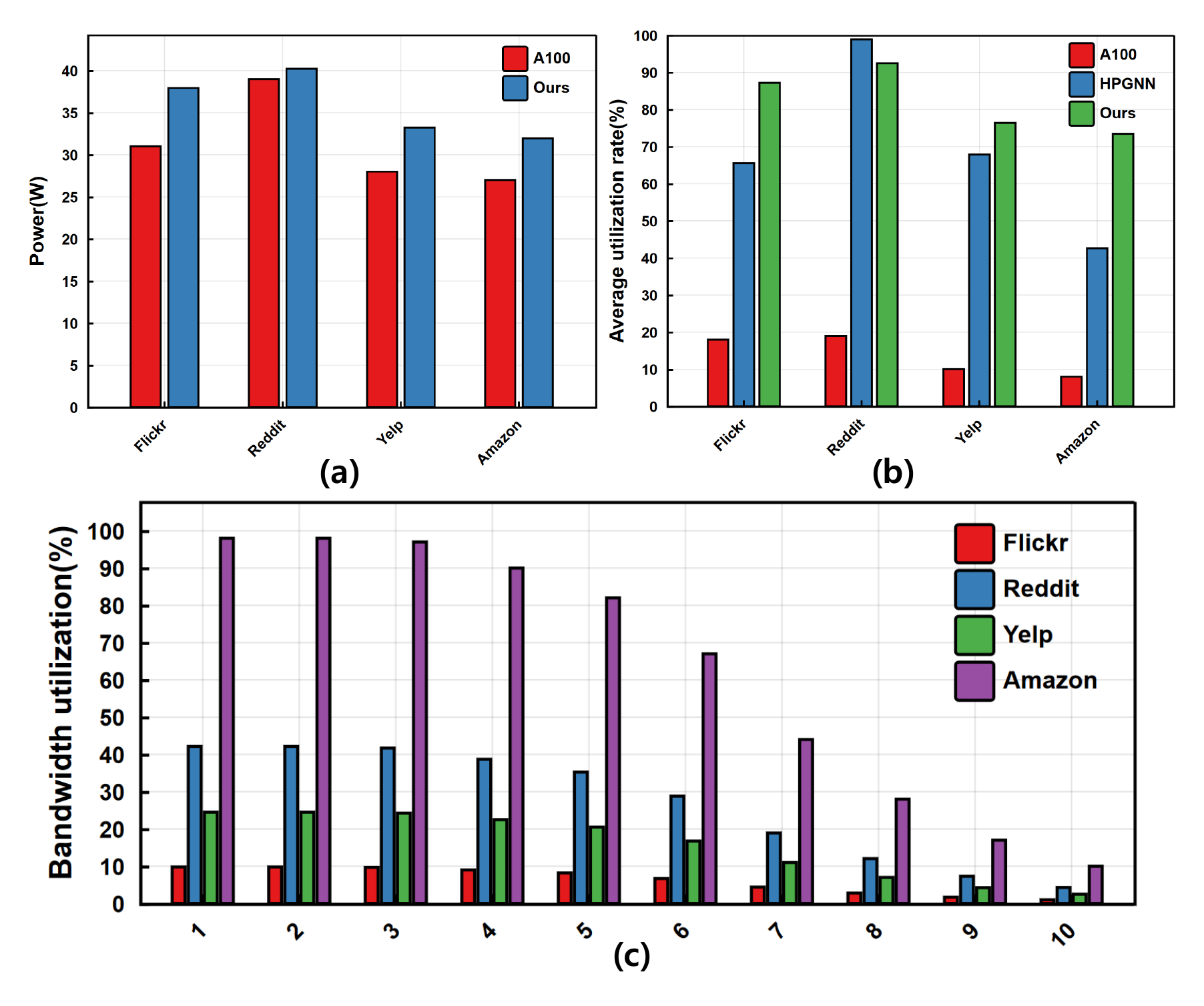}
  \caption{Power, Core and On-chip network utilization.}
  \label{test_2}
\end{figure}

In the case of multiple cores, due to the synchronization between each computational core (each core must wait for all cores to finish aggregation before starting combination), the actual time for a GCN layer in a multi-core setting is given by:
\begin{equation}
t_{multicore_{gcn}} = max(t_{singlecore_{gcn}})
\end{equation}
Therefore, from a single-core perspective, the average single-core utilization for Amazon and Yelp is higher than that for Reddit and Flickr. However, from a multi-core perspective, the opposite is true. As shown in Fig.\ref{test_2}(b), we tested the average utilization in a multi-core setting, and both Amazon and Yelp have lower utilization. This is because, in Fig.\ref{test}(c)(d), a small portion of cores have faster aggregation due to the power-law distribution of nodes (some nodes may have fewer neighbors), resulting in more cores waiting for the core with longer computation time, leading to a decrease in overall utilization. On the other hand, for Reddit, each computational core has a shorter average waiting time, resulting in higher utilization.

The bandwidth utilization of the interconnection network is generally related to the number of edges in the dataset (Fig.\ref{test_2}(c)). We conducted tests at 10 different time points during the aggregation stage, and the utilization gradually decreases as the aggregation progresses. This is also due to the uneven distribution of the number of node neighbors for each computational core.

\begin{figure} [htbp]
  \flushleft
  \includegraphics[scale=0.45]{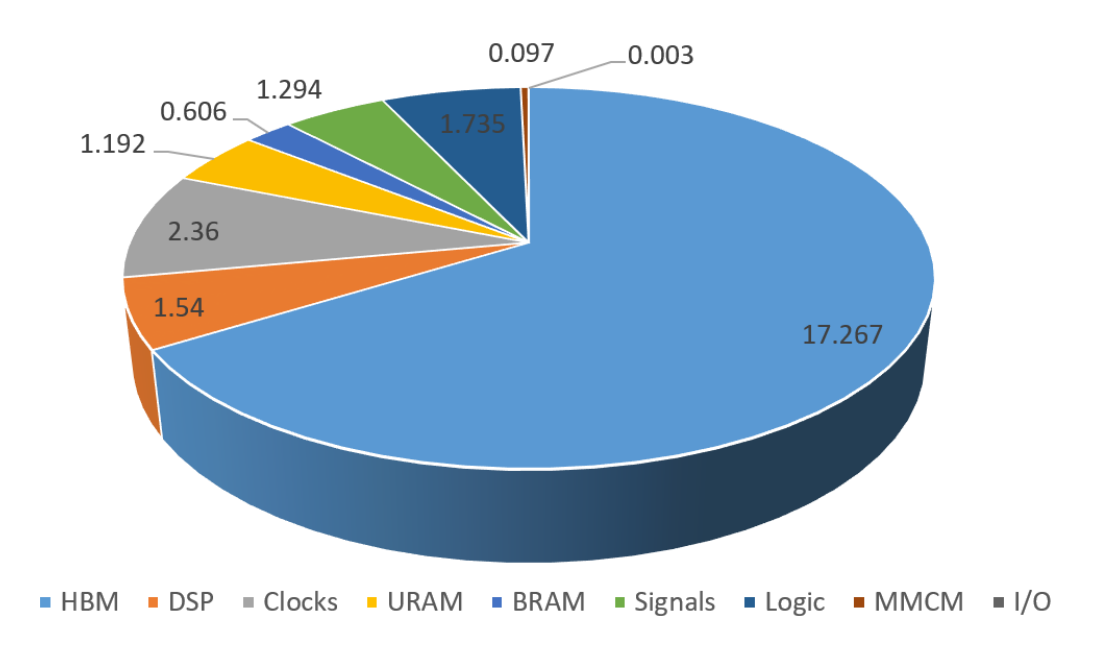}
  \caption{Dynamic On-chip Power.}
  \label{dpower}
\end{figure}
\subsubsection{Power Analysis}
We tested the power consumption of the entire accelerator board in Fig.\ref{test_2}(a), and the overall power consumption is higher compared to the A100 GPU. This may be due to two reasons: (1) The lower utilization of CudaCores in the GPU; (2) The NVIDIA A100 adopts a 7nm process while the Xilinx VCU128 uses a 16nm process. Since both use HBM, the power consumption is at a similar level compared to the GPU. We analyzed the dynamic power composition on the chip in detail (Fig.\ref{dpower}) and found that HBM accounts for 66.4$\%$ of the total on-chip power consumption, followed by Clock, DSP, Logic, and on-chip RAM. For deploying large-scale training tasks on FPGA, HBM is still necessary as it provides ample read-write bandwidth for DSP and helps alleviate some on-chip bandwidth pressure, which is crucial for improving performance.

\begin{table}[htbp]
  \centering
\scriptsize
  \caption{Comparison of time((s/epoch)) between GPU and other GNN training architecture, with 1024 batch size.}
    \label{compare}%
  \setlength{\tabcolsep}{2.5mm}{
  \renewcommand{\arraystretch}{1.5}
    \begin{tabular}{ccccccccccccccc}
    &&&&&&&&\\
    \midrule
     \midrule
    \textbf{}  & \multicolumn{3}{c}{\textbf{}} & \multicolumn{3}{c}{\textbf{GPU}} & \multicolumn{3}{c}{\textbf{HP-GNN\cite{lin2022hp}}} & \multicolumn{3}{c}{\textbf{Ours}} \\
    \midrule
     \midrule
    \multirow{3}*{\textbf{Platform}} & \multicolumn{3}{c}{Device} & \multicolumn{3}{c}{Nvidia A100} & \multicolumn{3}{c}{Alveo U250}  & \multicolumn{3}{c}{VCU128}  \\
    \cline{2-12}
 & \multicolumn{3}{c}{Peak Perf.} & \multicolumn{3}{c}{19.5TFLOPS} & \multicolumn{3}{c}{1.8TFLOPS}  & \multicolumn{3}{c}{2TFLOPS}  \\
    \cline{2-12}
 & \multicolumn{3}{c}{On-chip-Mem.}& \multicolumn{3}{c}{40MB} & \multicolumn{3}{c}{54MB} & \multicolumn{3}{c}{43MB}   \\
 
     \midrule
     \midrule
    \multirow{4}*{\textbf{NS-GCN}} & \multicolumn{3}{c}{Flickr} & \multicolumn{3}{c}{0.21($0.47\times$)}  & \multicolumn{3}{c}{0.16($1\times$)}  & \multicolumn{3}{c}{0.09($1.6\times$)}  \\
    \cline{2-12}
 & \multicolumn{3}{c}{Reddit} & \multicolumn{3}{c}{6.59($0.16\times$)} & \multicolumn{3}{c}{1.09($1\times$)}  & \multicolumn{3}{c}{1.05($1.03\times$)}   \\
    \cline{2-12}
 & \multicolumn{3}{c}{Yelp}& \multicolumn{3}{c}{2.90($0.38\times$)} & \multicolumn{3}{c}{1.35($1\times$)} & \multicolumn{3}{c}{1.11($1.22\times$)}   \\
    \cline{2-12}
 & \multicolumn{3}{c}{AmazonP.}& \multicolumn{3}{c}{5.06($0.38\times$)} & \multicolumn{3}{c}{3.49($1\times$)} & \multicolumn{3}{c}{1.92($1.81\times$)}    \\

     \midrule
     \midrule
    \multirow{4}*{\textbf{NS-SAGE}} & \multicolumn{3}{c}{Flickr} & \multicolumn{3}{c}{0.29($0.75\times$)}  & \multicolumn{3}{c}{0.22($1\times$)}  & \multicolumn{3}{c}{0.12($1.54\times$)}   \\
    \cline{2-12}
 & \multicolumn{3}{c}{Reddit} & \multicolumn{3}{c}{3.05($0.51\times$)} &\multicolumn{3}{c}{1.56($1\times$)}  & \multicolumn{3}{c}{1.37($1.13\times$)}   \\
    \cline{2-12}
 & \multicolumn{3}{c}{Yelp}& \multicolumn{3}{c}{3.51($0.52\times$)}&\multicolumn{3}{c}{1.85($1\times$)} & \multicolumn{3}{c}{1.64($1.12\times$)}   \\
    \cline{2-12}
 & \multicolumn{3}{c}{AmazonP.}& \multicolumn{3}{c}{6.83($0.70\times$)}&\multicolumn{3}{c}{4.83($1\times$)} & \multicolumn{3}{c}{3.65($1.32\times$)}   \\
     \midrule
    \bottomrule
    \end{tabular}}%
\end{table}%

\begin{table}[htbp]
  \centering
\scriptsize
  \caption{Resource Consumption.}
    \label{resources}%
  \setlength{\tabcolsep}{1.6mm}{
  \renewcommand{\arraystretch}{1.9}
    \begin{tabular}{ccccccccccccccccccccccccccccc}
    &&&&&&&&\\
    \midrule
     \midrule
    \textbf{Resources}  & \multicolumn{2}{c}{\textbf{LUTs}} & \multicolumn{2}{c}{\textbf{DSPs}} & \multicolumn{2}{c}{\textbf{FF}} & \multicolumn{2}{c}{\textbf{BRAM and URAM}} &
    \multicolumn{4}{c}{\textbf{HBM(GB)}}\\
    \midrule
     \midrule
    \textbf{Ours} & \multicolumn{2}{c}{807889} & \multicolumn{2}{c}{9000} & \multicolumn{2}{c}{1175200} & \multicolumn{2}{c}{24.5MB}  &  \multicolumn{1}{c}{1.8} &  \multicolumn{1}{c}{3.9} &  \multicolumn{1}{c}{2.5} &  \multicolumn{1}{c}{3.8} \\
        \textbf{HPGNN\cite{lin2022hp}} & \multicolumn{2}{c}{750960} & \multicolumn{2}{c}{8478} & \multicolumn{2}{c}{NA} & \multicolumn{2}{c}{16.2MB}  &  \multicolumn{4}{c}{NA}  \\
    \midrule
 \midrule

        \end{tabular}}%
\end{table}%

\subsection{Comparison with State-of-the-art}
At the architectural level, HPGNN\cite{lin2022hp} employs combination and aggregation separated computation engines, (Systolic Array, Scatter PE, and Gather PE). The Scatter PE and Gather PE are connected through a butterfly network. However, the separated computation engines can significantly impact performance when the computational workload is not balanced. During pipelined execution, the performance is limited by either the combination or aggregation engine. In datasets with a higher aggregation task load, the Scatter PE and Gather PE are busier than the Systolic Array, resulting in reduced utilization of the Systolic Array. In contrast, we adopt a unified aggregation and combination engine (Section 4.2) in our approach. The aggregation is initiated only when a certain number of neighboring nodes arrive at the target core. Therefore, the MAC utilization is only related to the transmission rate of the on-chip network. As a result, we can demonstrate higher resource utilization compared to HPGNN on datasets with uneven distribution (Fig.\ref{test_2}(b)), and get higher performance \ref{compare}.

On the other hand, in terms of on-chip network routing strategy, HPGNN does not provide specific control algorithms. In our work, we improved the hypercube network to deploy the algorithm and proposed an efficient routing algorithm for GNN on the hypercube to reduce message passing time.

Table \ref{resources} presents the basic on-chip resource consumption. Overall, our architecture consumes more LUTs and BRAM compared to HPGNN. This is because, in order to maximize HBM bandwidth, we deploy one DMA and its controller for every four channels, resulting in a total of eight DMAs, which is more than what DDR4 uses. In terms of storage consumption, we convert the edge table into a routing table, requiring more on-chip storage resources to store routing information. In terms of HBM overhead for training, Flickr, Reddit, Yelp, and Amazon consume approximately 1.8GB, 3.9GB, 2.5GB, and 3.8GB of storage, respectively. This is because we optimize the training data flow to reduce storage overhead (Table \ref{table_fx}), resulting in approximately one fewer edge table stored during training. In conclusion, we achieve high performance through data flow, route calculation, and storage optimization.


\bibliographystyle{ACM-Reference-Format}
\balance
\bibliography{sample-base}

\appendix

\end{document}